\documentclass[aps, pra, reprint, superscriptaddress]{revtex4-1}

\usepackage{amsmath}
\usepackage{amssymb}
\usepackage{wasysym}
\usepackage{graphicx}
\usepackage{hyperref}
\usepackage{color}
\usepackage{physics}
\usepackage{siunitx}
\usepackage{xcolor}
\usepackage{multirow}

\begin{document}

\title{Efficient conversion of closed-channel dominated Feshbach molecules \\of $^{23}$Na$^{40}$K to their absolute ground state}

\author{Roman~Bause}
\author{Akira~Kamijo}
\author{Xing-Yan~Chen}
\author{Marcel~Duda}
\author{Andreas~Schindewolf}
\affiliation{Max-Planck-Institut f\"{u}r Quantenoptik, 85748 Garching, Germany}
\affiliation{Munich Center for Quantum Science and Technology, 80799 M\"{u}nchen, Germany}
\author{Immanuel~Bloch}
\affiliation{Max-Planck-Institut f\"{u}r Quantenoptik, 85748 Garching, Germany}
\affiliation{Munich Center for Quantum Science and Technology, 80799 M\"{u}nchen, Germany}
\affiliation{Fakult\"{a}t f\"{u}r Physik, Ludwig-Maximilians-Universit\"{a}t, 80799 M\"{u}nchen, Germany}
\author{Xin-Yu~Luo}
\email{xinyu.luo@mpq.mpg.de}
\affiliation{Max-Planck-Institut f\"{u}r Quantenoptik, 85748 Garching, Germany}
\affiliation{Munich Center for Quantum Science and Technology, 80799 M\"{u}nchen, Germany}

\date{\today}

\begin{abstract}
We demonstrate the transfer of $^{23}$Na$^{40}$K molecules from a closed-channel dominated Feshbach-molecule state to the absolute ground state. The Feshbach molecules are initially created from a gas of sodium and potassium atoms via adiabatic ramping over a Feshbach resonance at \SI{78.3}{G}. The molecules are then transferred to the absolute ground state using stimulated Raman adiabatic passage with an intermediate state in the spin-orbit-coupled complex $|c^3 \Sigma^+, v=35, J=1 \rangle \sim |B^1\Pi, v=12, J=1\rangle$. Our measurements show that the pump transition dipole moment linearly increases with the closed-channel fraction. Thus, the pump-beam intensity can be two orders of magnitude lower than is necessary with open-channel dominated Feshbach molecules. We also demonstrate that the phase noise of the Raman lasers can be reduced by filter cavities, significantly improving the transfer efficiency.
\end{abstract}

\maketitle
\section{Introduction}
In recent years, ultracold molecules have emerged as a promising tool in quantum metrology, quantum simulation, and quantum computation~\cite{Bohn_2017, Safronova_2018, ACME_2018, Hughes_2020}. The field is now growing rapidly as evidenced by the recent cooling of multiple new molecular species~\cite{Prehn_2016, Anderegg_2017, Truppe_2017, Collopy_2018, Voges_2020, Cairncross_2021, Fitch_2021}. However, the creation of ultracold molecules is still a significant undertaking. Streamlining of experimental methods is therefore an important goal. Here, we present an improved and simplified method for the creation of ultracold $^{23}$Na$^{40}$K molecules in their absolute ground state.

Presently, the most controlled method for preparing samples of ultracold bialkali molecules is to first create a mixture of atoms and employ a Feshbach resonance to form molecules in a weakly bound spin-triplet state. These molecules exhibit only a small body-fixed dipole moment and are therefore typically transferred into the rovibrational ground state using stimulated Raman adiabatic passage (STIRAP)~\cite{Bergmann_1998, Vitanov_2017}. This method has been applied successfully to the species $^{40}$K$^{87}$Rb, $^{87}$Rb$^{133}$Cs, $^{23}$Na$^{40}$K, $^{23}$Na$^{87}$Rb, $^{23}$Na$^{6}$Li, and $^{23}$Na$^{39}$K~\cite{Ospelkaus_2008, Ni_2008, Debatin_2011, Takekoshi_2014, Molony_2014, Gregory_2015, Park_2015, Park_2015a, Guo_2016, Guo_2017, Rvachov_2017, Seesselberg_2018a, Liu_2019b, Voges_2019, Voges_2020}. The $^{23}$Na$^{40}$K molecule is a particularly interesting candidate for the creation of dipolar quantum gases due to its fermionic statistics and its comparatively large body-fixed dipole moment of \SI{2.7}{D}~\cite{Zuchowski_2010, Aymar_2005}.

STIRAP is still associated with considerable technical challenges as it requires two lasers with linewidths on the level of kilohertz or smaller, often at wavelengths not available with diode lasers. Furthermore, the transition dipole moments of the STIRAP transitions are small for some species, requiring high laser powers to reach adequate Rabi frequencies. For example, the first demonstration of STIRAP with $^{23}$Na$^{40}$K was carried out with a Ti:sapphire laser for the pump pulse at \SI{805}{\nano \meter} and a dye laser for the Stokes pulse at \SI{567}{\nano \meter}~\cite{Park_2015a}. A later implementation used tapered amplifiers seeded by diode lasers in combination with second-harmonic generation~\cite{Zhang_2019}. In our own previous work, we employed a different pathway with the pump pulse provided by a diode laser at \SI{488}{\nano\meter} and the Stokes pulse by a dye laser at \SI{652}{\nano\meter}~\cite{Seesselberg_2018a, Seesselberg_2018b, Bause_2020}. 

Our new method enables a transfer efficiency of 80\% using only two external-cavity diode lasers, an injection amplifier, and a single-pass second-harmonic generation module. This is possible due to an improved way of creating Feshbach molecules, allowing a larger overlap between the initial Feshbach-molecule state and the intermediate excited state. We also reduce off-resonant frequency components in the laser light by employing filter cavities with moderate finesse. As previously described in Refs.\ \cite{Akerman_2015b, Levine_2018}, laser noise is often a limiting factor for the high-fidelity addressing of quantum states. Laser-noise reduction is especially useful when using diode lasers for STIRAP, because their stabilization typically leads to strong noise contributions (servo bumps) at offset frequencies of $\sim \SI{1}{MHz}$. Since this is comparable to the desired Rabi frequencies for the STIRAP transitions, this noise is strongly detrimental. By employing filter cavities, we achieved a significant increase in STIRAP transfer efficiency compared to using unfiltered light.

\section{Association of Feshbach molecules}

\begin{figure}
\centering
\includegraphics{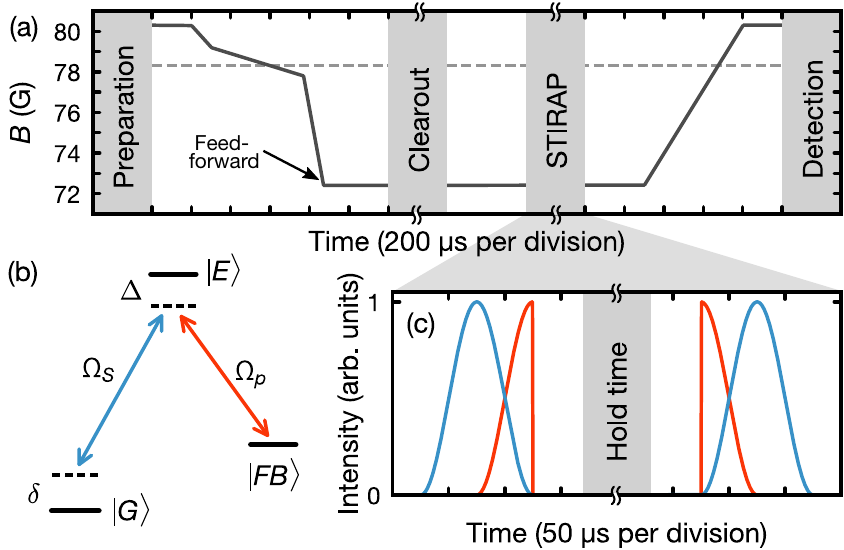}
\caption{Overview of the experimental sequence. (a)~Magnetic field ramps for Feshbach-molecule association and dissociation. ``Clearout'' stands for a 20-ms magnetic field gradient pulse, which removes unassociated atoms. The dashed gray line indicates the position of the Feshbach resonance. (b)~Three-level system used for STIRAP. The intermediate excited state $|E\rangle$ is connected to the Feshbach-molecule state $|FB\rangle$ via the pump beam (bright red) and to the ground state $|G\rangle$ via the Stokes beam (dark blue). (c)~STIRAP pulse shapes for the pump and Stokes beams. A pulse duration of $\SI{50}{\mu s}$ is assumed here. The typical peak intensities of the beams are $\SI{200}{W/cm^2}$ for the pump beam and $\SI{0.3}{W/cm^2}$ for the Stokes beam. This corresponds to powers of \SI{10}{mW} and $\SI{15}{\mu W}$, respectively. The length of the hold time depends on the specific experiment that is performed.}
\label{fig-sequence}
\end{figure}

\begin{figure}
\centering
\includegraphics{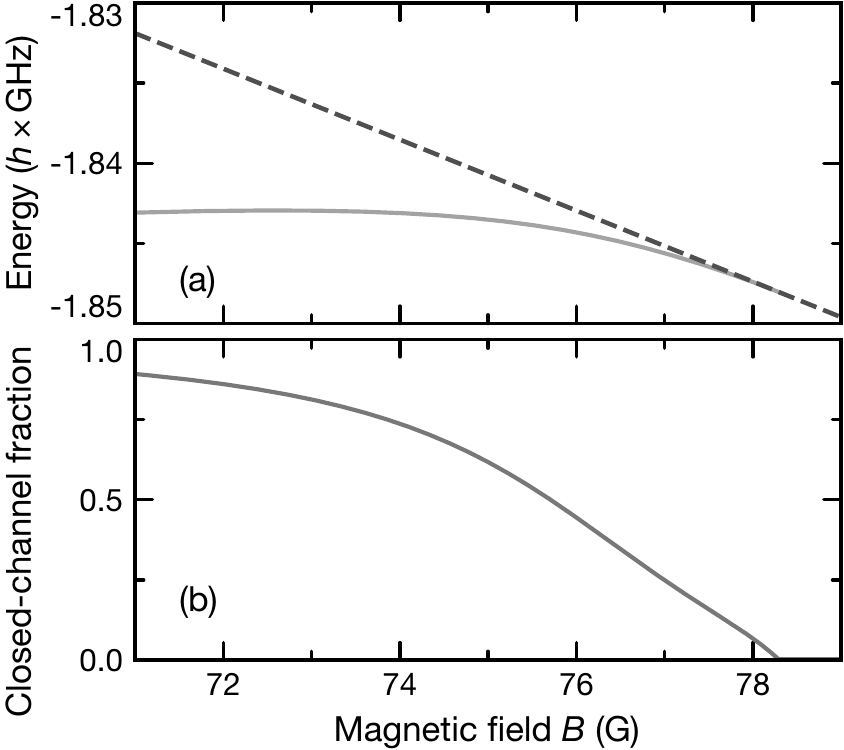}
\caption{Properties of atoms and Feshbach molecules depending on the magnetic field. (a)~Zeeman shift of atoms and Feshbach molecules. The dashed line shows the energy of an unbound atom pair, relative to the hyperfine center of mass. The solid line branches off at the resonance position and shows the energy of the Feshbach molecule. At $B=\SI{72.4}{G}$, there is an extremum where the energy becomes first-order insensitive to field changes. (b)~Closed-channel fraction of Feshbach molecules depending on the magnetic field.}
\label{fig-ccf}
\end{figure}

We begin our experiments by preparing a gas of $1 \times 10^5$ sodium atoms in the state $|F=1, m_F=1\rangle$ and $2 \times 10^5$ potassium atoms in $|F=9/2, m_F=-9/2\rangle$, where $F$ denotes the total atomic angular momentum and $m_F$ its projection onto the magnetic field direction. The atomic mixture is prepared in a crossed-beam optical dipole trap with trap frequencies of $2\pi \times$(50, 72, 191)$\,$Hz for Na and $2\pi \times$(54, 88, 223)$\,$Hz for K in the $(x,y,z)$ directions, respectively, where the $z$-axis points along the direction of gravity. At the end of the preparation, we reach a  temperature of about \SI{300}{nK} at a magnetic field of \SI{80.3}{G}, where the inter-species scattering length vanishes. We then ramp the magnetic field over a Feshbach resonance at \SI{78.3}{G} in three steps to create about $4 \times 10^4$ weakly bound Feshbach molecules, corresponding to a 40\% transfer efficiency of the minority species. The association procedure is shown in Fig.~\ref{fig-sequence}(a). In the first part, we quickly ramp the field to \SI{79.2}{G}, slightly above the resonance, over $\SI{100}{\mu s}$. The second ramp, where the actual association of Feshbach molecules happens, changes the field to \SI{77.8}{G} within $\SI{470}{\mu s}$. The duration and the end point of this ramp are chosen to optimize the association efficiency. However, at \SI{77.8}{G}, the molecules are highly unstable against collisions with residual atoms, so we perform a third ramp to a final field of \SI{72.4}{G} within $\SI{100}{\mu s}$ to minimize inelastic atom-molecule collisions. The final value is chosen because the following STIRAP transfer is least sensitive to field fluctuations there, see Fig.\ \ref{fig-ccf}(a). To achieve optimal stability for STIRAP and following experiments, the final ramp is performed with a feed-forward procedure, which is designed to minimize the effect of eddy currents in conductive parts of the apparatus close to the molecule sample. This is achieved by changing the current through the magnetic-field coil in such a way that the known eddy currents are compensated, following the idea described in Ref.~\cite{Olsen_2008}. Finally, we apply a magnetic-field gradient of \SI{40}{G/cm} for a duration of \SI{20}{ms}, which removes residual atoms from the dipole trap while leaving the Feshbach molecules unaffected due to their lack of magnetic moment.

This association procedure is similar to the ones used with other molecules such as $^{40}$K$^{87}$Rb~\cite{Ni_2008}, however it has previously not been applied to $^{23}$Na$^{40}$K. Instead, previous work on this species has relied on radiofrequency association. This was limited to 10-15\% efficiency~\cite{Wu_2012, Seesselberg_2018a} because radiofrequency association is essentially instantaneous compared to the timescale of collisions in the gas. The transfer efficiency is then determined by the phase-space overlap of the two gases at a weak interspecies interaction. In contrast, an adiabatic ramp of the magnetic field across the Feshbach resonance allows multiple elastic collisions during the association, thereby effectively increasing the phase-space overlap and the transfer efficiency. 

The Feshbach resonance at \SI{78.3}{G} is an ideal choice for our association procedure due to its width of \SI{5.3}{G}, much narrower than the previously used resonance at \SI{89.7}{G}, which is \SI{9.3}{G} wide~\cite{Chen_2021}. Because the magnetic-field ramp rate is limited by the speed of the feed-forward procedure, a smaller ramp range results in fewer inelastic collisions during the ramp. In addition, on the right side of the 78.3-G resonance, the interspecies interaction can be tuned to be moderately repulsive, which is better suited for sympathetic cooling than the stronger attractive interaction on the right side of the 89.7-G resonance.

The Feshbach-molecule state $|FB\rangle$ can be described as a superposition of the open-channel part, which corresponds to two unbound atoms, and the closed-channel part, which corresponds to the least-bound vibrational level of the electronic $a^3\Sigma$ manifold. As shown in Fig.~\ref{fig-ccf}(b), the relative weight of the open- and closed-channel contributions depend on the magnetic field, with the closed-channel fraction becoming larger for the more deeply-bound Feshbach molecules far below the resonance.

\section{STIRAP setup}
\label{section-setup}
\begin{figure*}
\centering
\includegraphics{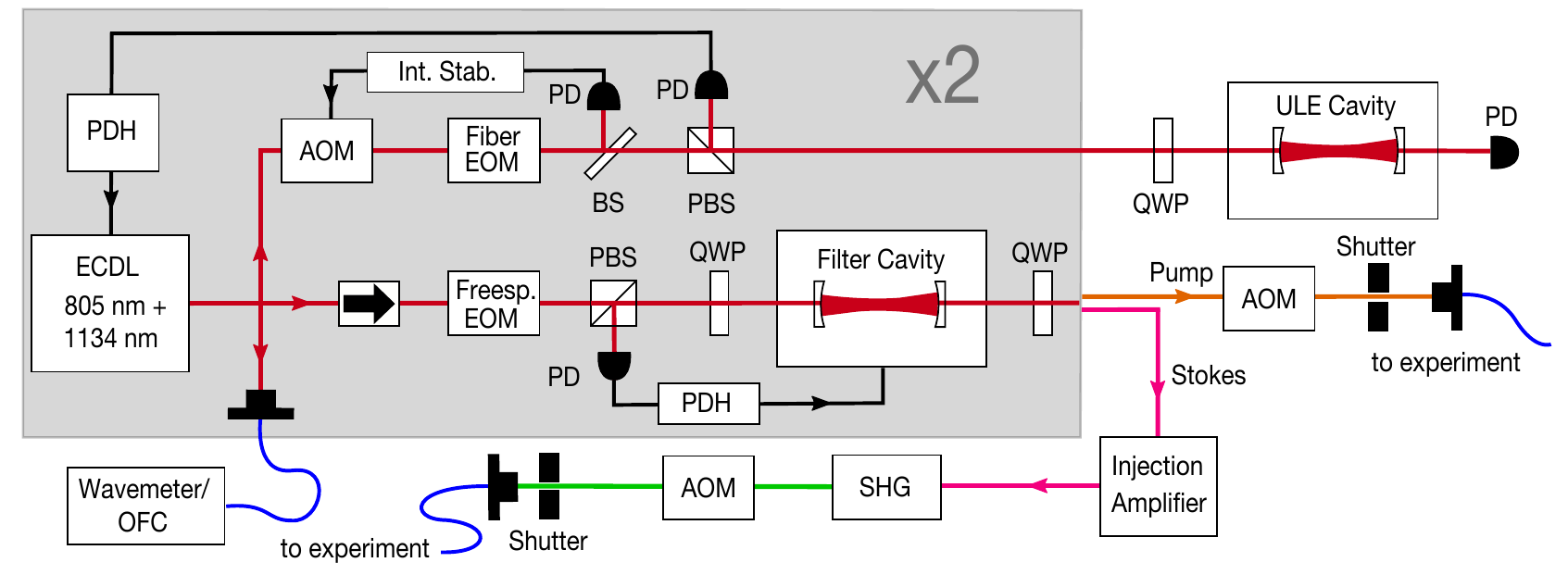}
\caption{Schematic of the setup. The part of the setup inside the gray area exists twice, once for the pump and once for the Stokes beam. Laser light is generated by two external-cavity diode lasers (ECDL), and part of the power is split off to be coupled into the ULE cavity. Photodiodes (PD) and a PI-loop controlling acousto-optical modulators (AOM) are used for intensity stabilization. Sidebands for the PDH lock are generated using electro-optical modulators (EOM). The error signal for the lock is created from the cavity reflection using a quarter-waveplate (QWP) and a polarizing beamsplitter (PBS). Another PD can be used to monitor the ULE cavity transmission. The main parts of the light of each laser are coupled through filter cavities, which are PDH-locked to the laser frequency. The light for the Stokes beam is then injection-amplified and frequency-doubled. Both beams go through AOMs to create the STIRAP pulse shape and are subsequently fiber-coupled to be sent to the molecules. The frequency of both beams can be monitored using a wavemeter or a beat note with an optical frequency comb (OFC).}
\label{fig-setup}
\end{figure*}

The STIRAP pathway relies on an intermediate excited state $|E\rangle$ in the complex $|c^3\Sigma^+, v=35, J=1\rangle \sim |B^1 \Pi, v=12, J=1\rangle$, the same which was previously used in~\cite{Park_2015a, Liu_2019b}. This state features strong mixing between triplet (64\%) and singlet (36\%) states~\cite{Park_2015}, making it ideal for transfer from the triplet Feshbach-molecule state to the singlet ground state $|G\rangle = |X^1\Sigma^+, v=0, J=0\rangle$. Here, $v$ stands for the vibrational quantum number and $J$ for the total angular momentum excluding nuclear spin. The contribution of different nuclear spin states to the intermediate state is reported in Table \ref{table-excited-state}.

To address this intermediate state, we generate light at \SI{567}{\nano\meter} and \SI{805}{\nano\meter} with the setup shown in Fig.~\ref{fig-setup}. The setup is based on two external-cavity diode lasers (DLPro, Toptica) at wavelengths of \SI{1134}{\nano\meter} and \SI{805}{\nano\meter}, and an ultrastable, dual-wavelength optical cavity (Stable Laser Systems) of finesse $\mathcal{F}_{1134} = 3.530(6) \times 10^4$ and $\mathcal{F}_{805} = 4.58(12) \times 10^4$ at these wavelengths, respectively. The reference cavity spacer is made of ultra-low-expansion (ULE) material and is mounted inside a vacuum chamber at a pressure of \SI{1.3e-6}{mbar}. The chamber is designed to damp vibration and reduce thermal contact with the environment, such that the cavity spacer can be kept within \SI{0.01}{\kelvin} around its zero-expansion temperature of \SI{30.74(10)}{\celsius}. This is achieved with in-vacuum temperature stabilization and a two-layer copper shield, which allows us to reach an exponential time constant of 6.5 hours for temperature changes of the system.

We employ fiber-coupled EOMs (PM830 and PM1170, Jenoptik) to modulate sidebands at frequencies of up to \SI{750}{MHz} onto each laser beam and subsequently lock these sidebands to the reference cavity using the Pound-Drever-Hall (PDH) locking technique~\cite{Drever_1983}. In combination with the cavity's free spectral range of \SI{1.5}{GHz}, this allows us to stabilize the lasers to any frequency within their operation range. The fiber EOMs also exhibit very low residual amplitude modulation compared to bulk EOMs. We stabilize the input laser powers of the ULE cavity to minimize its thermal drift due to light absorption on the cavity mirrors. With the PDH lock, we achieve a long-term stability limited by the cavity spacer's aging process, which initially causes a linear frequency drift of the order of $2 \pi \times \SI{5}{kHz}$ per day. The drift slows down exponentially with a measured time constant of 375 days as the spacer approaches equilibrium~\cite{Sterr_2009, Ito_2017}.

\begin{table}
	\caption{Quantum numbers of the STIRAP intermediate state, $|E\rangle$. The projections of the nuclear spin of the atoms are labeled $m_{I,\mathrm{Na}}$ and $m_{I,\mathrm{K}}$. Only the eight strongest contributions, which together make up 99.2\% of the weight, are shown.}
	\label{table-excited-state}
	\renewcommand{\arraystretch}{1.3}
	\begin{tabular}{c | c c c l}
		\hline
		Manifold & \hspace{0.2cm} $m_J$ \hspace{0.2cm} & \hspace{0.2cm} $m_{I,\mathrm{Na}}$ \hspace{0.2cm} & \hspace{0.2cm} $m_{I,\mathrm{K}}$ \hspace{0.2cm} & Weight \\ \hline
		& $0$ & $1/2$ & $-3$ & $0.0145$ \\
		\multirow{2}{*}{$B^1\Pi$}
		& $0$ & $3/2$ & $-4$ & $0.0754$ \\
		& $1$ & $-1/2$ & $-3$ & $0.0098$ \\
		& $1$ & $1/2$ & $-4$ & $0.258$ \\ \hline
		& $0$ & $1/2$ & $-3$ & $0.0257$ \\
		\multirow{2}{*}{$c^3\Sigma^+$}
		& $0$ & $3/2$ & $-4$ & $0.133$ \\
		& $1$ & $-1/2$ & $-3$ & $0.0173$ \\
		& $1$ & $1/2$ & $-4$ & $0.458$ \\
	\end{tabular}
\end{table}

\begin{figure}
\centering
\includegraphics{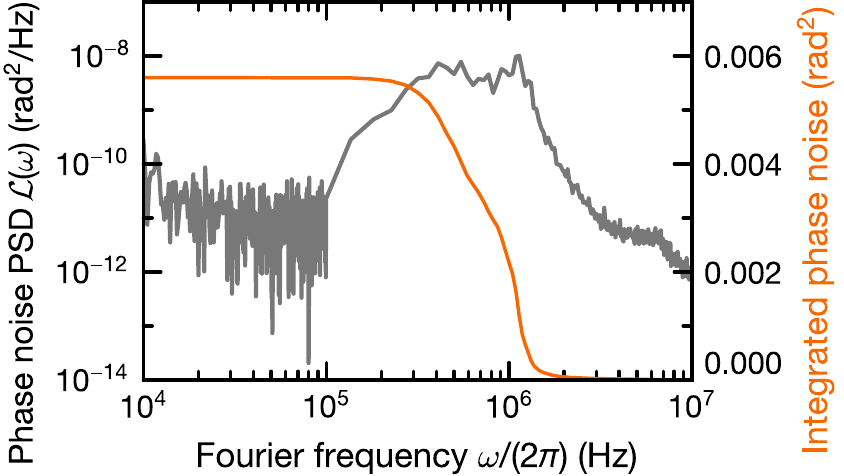}
\caption{Power spectral density (PSD) of the phase noise of the pump laser, $\mathcal{L(\omega)}$, relative to the ULE reference cavity. This was determined from the measured power spectral density of the PDH error signal. The PSD is shown in gray. The orange line corresponds to the integrated PSD, with the integration beginning at high frequencies.}
\label{fig-noise-spectral-density}
\end{figure}

We characterize the quality of the laser lock by the power spectrum of the in-loop PDH error signal. From this, the phase noise power spectral density $\mathcal{L}(\omega)$ can be determined as described in Ref.~\cite{Schmid_2019}, see Fig.~\ref{fig-noise-spectral-density}. At sufficiently high frequencies and small modulation, the integral of $\mathcal{L}(\omega)$ is equal to the total power in the phase noise pedestal $P_\phi$. The servo bumps of the PDH lock are at frequencies of $2 \pi \times \SI{1.2}{\mega\hertz}$. The integrated power of the laser's phase noise pedestal is almost exclusively in these servo bumps and is about 0.6\% of the total laser power.

As described in section \ref{section-stirap-efficiency}, the servo bumps are in a frequency range which is especially problematic for STIRAP efficiency. Hence we reduce the power in the servo bumps using two optical filter cavities with finesse $\mathcal{F} = 4430(14)$ and $\mathcal{F} = 5110(30)$ for the 1134-nm and the 805-nm light, respectively. Using piezo-driven mirrors, the filter cavities are PDH-locked to achieve optimal transmission. Sending the light through these cavities is expected to reduce the servo bumps by \SI{22}{dB}. A Faraday isolator is used before the filter cavity input to block detrimental back-reflection into the ULE-cavity setup. The filter-cavity setup reduces the available power of both STIRAP beams by about 50\%, mostly due to imperfect coupling into the cavities.

The 1134-nm light is then injection-amplified to a total power of \SI{105}{mW} using an anti-reflection-coated gain chip (GC-1180-100-TO-200-B, Innolume), and frequency-doubled in a single-pass periodically-poled lithium-niobate waveguide (HC Photonics) to create \SI{3}{mW} of 567-nm light. Both beams are sent to the experimental chamber through shutters and acousto-optical modulators, which allow us to create the desired pulse shape for STIRAP. As previously discussed in Refs.~\cite{Seesselberg_2018a, Yatsenko_2002, Yatsenko_2014}, we use a pulse shape given by
\begin{equation}
I(t) =
\begin{cases}
I_0 \cos^2\left(\frac{\pi t}{2\tau}\right), \quad -\tau < t < \tau \\
0 \qquad \qquad \qquad \text{otherwise}
\end{cases}
\end{equation}
with the peak intensity $I_0$, see Fig.~\ref{fig-sequence}(b). The beam intensities are independently chosen to achieve comparable Rabi frequencies of around $2 \pi \times \SI{2.5}{MHz}$ for each beam. For the transfer to the ground state, the Stokes pulse precedes the pump pulse by $\tau$, and for the reverse transfer they are interchanged. For both pump pulses, one half is cut off, which simplifies our experimental control with no impact on the STIRAP itself. 

\section{Results}
\subsection{Spectroscopy of the pump and Stokes transitions}
We performed photoassociation spectroscopy on the Feshbach molecules to identify our target intermediate state as previously described in the work of Park \textit{et al.~}\cite{Park_2015}. We found the STIRAP intermediate state at a pump laser frequency of $2 \pi \times \SI{372554391(1)}{MHz}$, as determined using a beat note with an optical frequency comb. In the following, the pump laser frequency is given as a detuning $\Delta_p$ relative to this transition.

\begin{figure}
\centering
\includegraphics{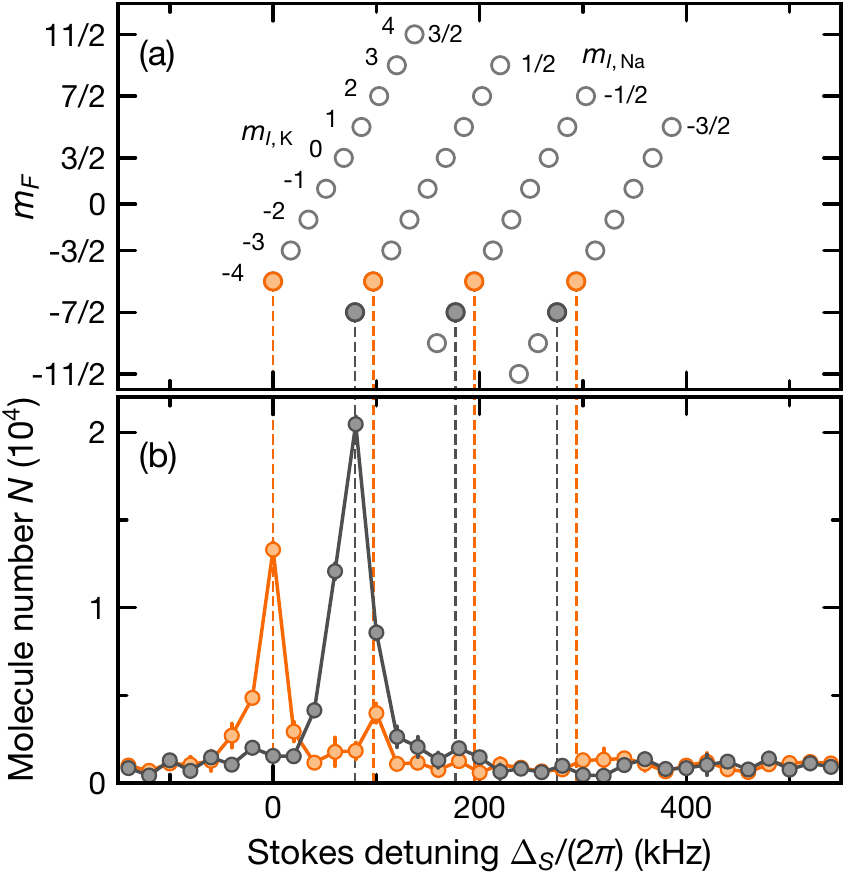}
\caption{Two-photon spectroscopy of the ground-state hyperfine structure. (a)~Hyperfine sublevels of the rovibrational ground state. States with $m_F = -5/2$ can be reached from $|E\rangle$ via $\pi$-transitions and are shown in bright orange. States with $m_F = -7/2$ can be reached via $\sigma^-$-transitions and are shown in dark gray. (b)~Dark resonance spectra. Data for parallel polarization of the Stokes beam, corresponding to $\pi$ transitions is shown in orange, data for orthogonal polarization, corresponding to $\sigma$ transitions, in gray. Frequencies are given relative to the transition from the STIRAP excited state to the absolute ground state, which corresponds to $2 \pi \times \SI{528805718}{MHz}$.}
\label{fig-gs-hyperfine-structure}
\end{figure}

To find the correct frequency for the Stokes beam, we performed dark-state spectroscopy of the ground state. For this, the Feshbach molecules were illuminated with both the pump and the Stokes beam simultaneously for $\SI{200}{\mu s}$. The intensities were $\SI{5}{W/cm^2}$ and $\SI{4}{mW/cm^2}$ for the pump and Stokes beam, respectively, at a beam waist of $\SI{30}{\mu m} \times \SI{100}{\mu m}$ for both beams. The pump light was kept on resonance with the intermediate state while the Stokes beam frequency was scanned. With the Stokes beam away from resonance, the pump beam power is sufficient to photoassociate all Feshbach molecules during the illumination time. However, when the Stokes beam becomes resonant, the light shift of the excited state prevents photoassociation, thus showing up as increased remaining molecule numbers. We scanned the hyperfine structure of the rovibrational ground state for both horizontal and vertical polarization of the Stokes beam and assigned the observed peaks to find the absolute ground state (see Fig.\ \ref{fig-gs-hyperfine-structure}). This state has $m_F = -5/2$ character and we determined the Stokes transition frequency to be $2 \pi \times \SI{528805718(1)}{MHz}$. The detuning of the Stokes laser from this frequency is denoted by $\Delta_S$. As the pump frequency remained the same throughout the measurement, the two-photon detuning $\delta = \Delta_S - \Delta_p$ was changed together with the Stokes detuning, and the one-photon detuning $\Delta = \Delta_p$ was always zero. In combination with the photoassociation data, these measurements yielded a binding energy of the absolute ground state of $\SI{5212.04443(3)}{cm^{-1}}$, relative to the hyperfine center of mass of the free atoms, consistent with the value found in Ref.~\cite{Park_2015}.

\subsection{Determination of Rabi frequencies}
Next, we performed two-photon spectroscopy on Feshbach-molecule samples  to determine the Rabi frequencies associated with the two transitions. To obtain the spectra, Feshbach molecules were illuminated with the Stokes beam at a high intensity of around $\SI{50}{W/cm^2}$. Its frequency was kept close to the previously determined value for the transition, i.e.\ $\Delta_S \approx 0$. Hence, for this measurement, the condition $\delta = \Delta$ was always fulfilled. While the Stokes beam remained on, the pump beam was added for $\SI{50}{\mu s}$ at various detunings. Note that we kept the convention of naming the transitions ``pump'' and ``Stokes'', as is typical for STIRAP, even though in the context of the measurements in this section, they are usually called ``probe'' and ``coupling'', respectively~\cite{Fleischhauer_2005}. After the illumination, we measured the number of remaining molecules. The resulting spectrum features a dark state of the system where no molecules are lost when both lasers are on resonance, i.e.\ $\Delta = 0$. Tuning $\Delta$ reveals two dips, which are split by the Stokes laser Rabi frequency $\Omega_S$. The shape of the entire feature is described by~\cite{Fleischhauer_2005, Seesselberg_2018a}
\begin{equation}
\label{eq-eit-shape}
N = N_0 \exp \left( -t \Omega_p^2 \frac{4 \gamma \delta^2}{|\Omega_S^2 + 2i\delta (\gamma + 2i\Delta)|^2}\right),
\end{equation}
where $N$ and $N_0$ are the remaining and initial molecule numbers, respectively, $\gamma$ is the linewidth of the intermediate state $|E\rangle$, and $t$ is the duration of the pump pulse. Since in our case $\gamma \approx \Omega_S$, we are in an intermediate regime between electromagnetically induced transparency (EIT), which results purely from interference, and Autler-Townes splitting, which results from a light shift due to the strong Stokes beam~\cite{Fleischhauer_2005}.

\begin{figure}
\centering
\includegraphics{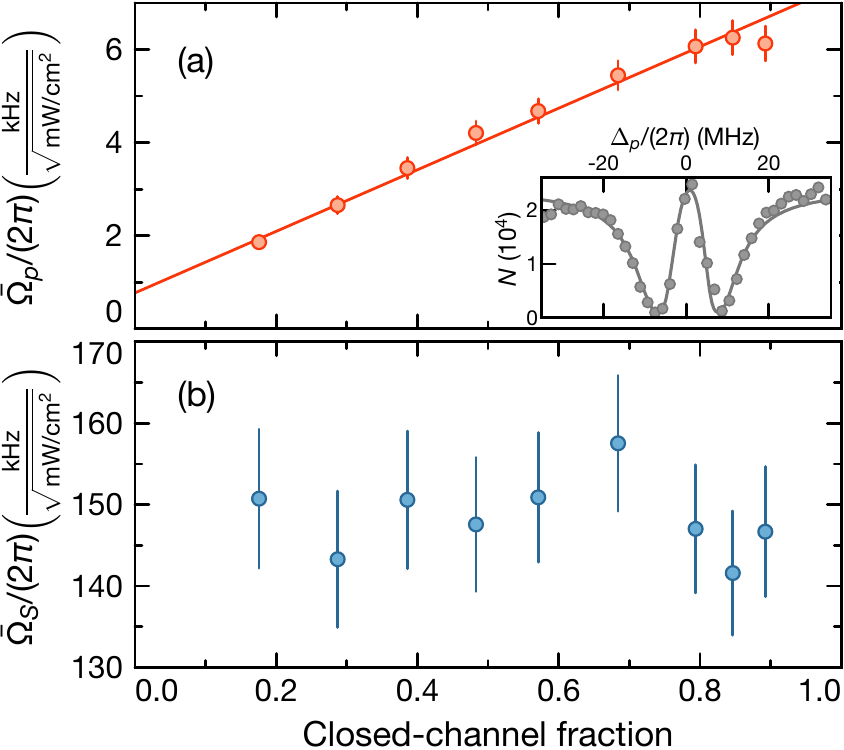}
\caption{Normalized Rabi frequencies depending on closed-channel fraction as determined from EIT spectra. (a)~Normalized pump Rabi frequency $\bar{\Omega}_p$. The solid line is a linear fit, which yields a slope of $2 \pi \times 6.6 \, \mathrm{kHz/ \sqrt{mW/cm ^2}}$ and an offset of $2 \pi \times 0.8 \, \mathrm{kHz/ \sqrt{mW/cm ^2}}$. Inset: Two-photon spectrum used to determine Rabi frequencies with $\Delta_S \approx 0$ at $B=\SI{72.4}{G}$. The solid line is a fit of Eq.~(\ref{eq-eit-shape}), which yields $\Omega_p = 2 \pi \times \SI{0.29(1)}{MHz}$,  $\Omega_S = 2 \pi \times \SI{17.1(3)}{MHz}$, and $\gamma = 2 \pi \times \SI{11(1)}{MHz}$. (b)~Normalized Stokes Rabi frequency $\bar{\Omega}_S$. Error bars represent the $1\sigma$ uncertainty of the fit. }
\label{fig-normalized-rabi-frequencies}
\end{figure}

We then prepared Feshbach molecules at different magnetic background fields by performing an additional magnetic field ramp after the completion of the clearout. This allowed us to vary the closed-channel fraction of molecules and investigate its effect on the transition dipole moments of the pump and Stokes transitions. To this end, we determined the normalized Rabi frequencies $\bar{\Omega}_p = \Omega_p / \sqrt{I_p}$ and $\bar{\Omega}_S = \Omega_S / \sqrt{I_S}$ by fitting Eq.~(\ref{eq-eit-shape}) to the obtained spectra. The normalized Rabi frequencies are directly proportional to the respective transition dipole moments. Here, $I_p$ $(I_S)$ stands for the intensity of the pump (Stokes) beam. As the results in Fig.~\ref{fig-normalized-rabi-frequencies} show, $\bar{\Omega}_p$ is approximately proportional to the closed-channel fraction. This is expected because only the closed-channel part of the Feshbach-molecule state couples with the excited state.

For near-zero closed-channel fractions, we found $\bar{\Omega}_p = 2 \pi \times 0.8 \, \mathrm{kHz/ \sqrt{mW/cm ^2}}$, which is comparable to the value previously reported in Ref.~\cite{Park_2015} using Feshbach molecules created via radiofrequency association. In contrast, $\bar{\Omega}_S$ remained essentially constant, as the states $|E\rangle$ and $|G\rangle$ are only weakly affected by the magnetic field. At $B = \SI{72.4}{G}$, which corresponds to a closed-channel fraction of 0.84, we found $\bar{\Omega}_p = 2 \pi \times 6.3(4) \, \mathrm{kHz/ \sqrt{mW/cm ^2}}$ and $\bar{\Omega}_S = 2 \pi \times 142(8) \, \mathrm{kHz/ \sqrt{mW/cm ^2}}$. The measured excited-state linewidth is $\gamma = 2 \pi \times \SI{11(1)}{MHz}$. Thus, the transition dipole moment of the pump transition at $B=\SI{72.4}{G}$ is 14 times larger than was demonstrated in Ref.~\cite{Park_2015} with a closed-channel fraction of a few percent. This means the same STIRAP efficiency can be achieved with two orders of magnitude lower laser intensity, making amplification of the pump laser unnecessary. The summary of the performance achieved with previous STIRAP setups shown in Table~\ref{table-stirap-overview} provides some context to this result.

\begin{table*}
	\caption{Overview of experimentally investigated STIRAP schemes from Feshbach states to $X^1\Sigma^+$ rovibrational ground states of bialkali molecules. $1 \, \mathrm{kHz/ \sqrt{mW/cm^2}}$ corresponds to a transition dipole moment of $9.00 \times 10^{-7} \, ea_0$. The pump and Stokes laser wavelengths are denoted by $\lambda_p$ and $\lambda_S$, respectively. Note that STIRAP has not yet been demonstrated for some of the listed transition pairs.}
	\label{table-stirap-overview}
	\renewcommand{\arraystretch}{1.3}
	\begin{tabular}{l l r r r r r c l}
		\hline
		Molecule \rule{0pt}{4ex} \hspace{1em} & Interm.\ state & \hspace{0.3em} $\lambda_p$ (nm) & \hspace{0.3em} $\lambda_S$ (nm) & \hspace{0.5em} ${\displaystyle\frac{\bar{\Omega}_p}{2\pi}} \,\left(\mathrm{\frac{kHz}{\sqrt{mW/cm^2}}} \right)$ & \hspace{0.5em} ${\displaystyle\frac{\bar{\Omega}_S}{2\pi}} \, \left(\mathrm{\frac{kHz}{\sqrt{mW/cm^2}}} \right)$ &  \hspace{0.7em} ${\displaystyle\frac{\gamma}{2\pi}}\,(\mathrm{MHz})$ &  \hspace{1.5em} $\eta$  \hspace{1.5em} & Sources \\ \hline
		$^6$Li$^{40}$K & $A^1 \Sigma^+$\footnote{A Feshbach state with a strong singlet contribution was used, avoiding the need for singlet-triplet mixing in the intermediate state.} & 1119 & 665 & 2.7(2) & 12(2) & 5 & -- & \cite{Yang_2020} \\
		$^{23}$Na$^{39}$K & $c^3 \Sigma ^+ \sim B^1 \Pi$ & 816 & 572 & -- & 65.2 & 6 & 70\% & \cite{Voges_2019, Voges_2020} \\
		$^{23}$Na$^{40}$K & $c^3 \Sigma ^+ \sim B^1 \Pi$ & 805 & 567 & 0.45(10) & 25(5) & 9(1) & 78\% & \cite{Park_2015, Park_2015a, Liu_2019b} \\
		$^{23}$Na$^{40}$K & $d^3\Pi \sim D^1 \Pi$ & 652 & 487 & 0.6 & 6 & 20 & 50\% & \cite{Seesselberg_2018a} \\
		$^{23}$Na$^{40}$K & $c^3 \Sigma ^+ \sim B^1 \Pi$ & 805 & 567 & 6.3(4) & 142(8) & 11(1) & 80\% & This work \\
		$^{23}$Na$^{87}$Rb & $b^3\Pi \sim A^1 \Sigma^+$ & 1248 & 769 & 1.01 & 23.2 & 0.67 & 93\% & \cite{Guo_2016, Guo_2017, Ye_2018} \\
		$^{23}$Na$^{87}$Rb & $B^1 \Pi \sim c^3\Sigma^+$ & 803 & 574 & 3.4 & 100 & 6 & -- & \cite{Guo_2018c} \\
		$^{23}$Na$^{133}$Cs & $c^3\Sigma^+ \sim B^1 \Pi$ & 922 & 635 & 10.1(13) & 257(3) & 120 & 82\%\footnote{This efficiency was achieved with off-resonant Raman scattering rather than with STIRAP.} & \cite{Cairncross_2021} \\
		$^{40}$K$^{87}$Rb & $2^3\Sigma^+ \sim 1^1\Pi$ & 968 & 689 & 6(2) & 13(3)& -- & 89\% & \cite{Ni_2008, Ni_2010b, Moses_2015} \\
		$^{87}$Rb$^{133}$Cs & $b^3\Pi \sim A^1 \Sigma^+$ & 1557 & 977 & 0.84(24) & 2.76(67) & 0.135(10)\footnote{Ref.~\cite{Molony_2016} reported a different value of $\gamma = 2\pi \times \SI{35(3)}{kHz}.$} & 92\% & \cite{Takekoshi_2014, Molony_2016} \\ \hline
	\end{tabular}
\end{table*}

\subsection{STIRAP and laser noise reduction}
\label{section-stirap-efficiency}

\begin{figure}
\centering
\includegraphics{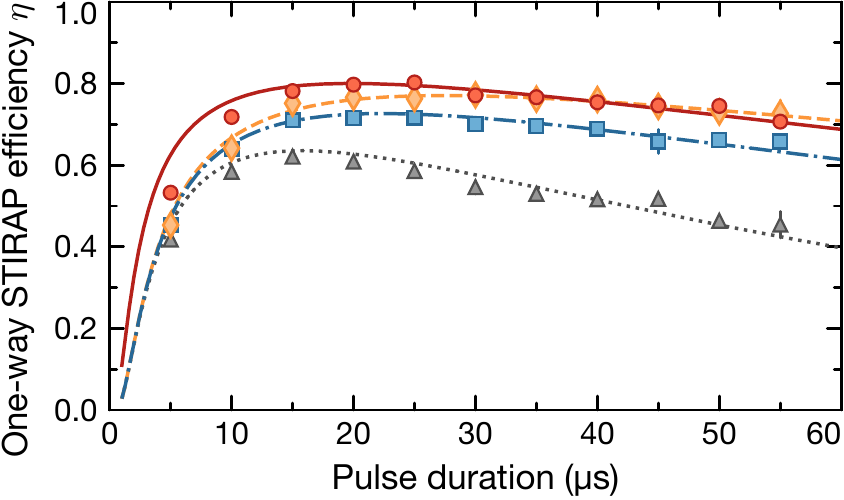}
\caption{One-way STIRAP efficiency depending on pulse duration, Rabi frequencies, and presence of filter cavities. The lines are fits of Eq.\ (\ref{eq-stirap-efficiency}) to the data. Grey triangles, corresponding to the dotted line, were taken with no filter cavities. Orange diamonds, corresponding to the dashed line, with both filter cavities. Blue squares, corresponding to the dash-dotted line, were taken with only the pump filter cavity. In these three datasets, the Rabi frequencies were $\Omega_p = \Omega_S = 2 \pi \times \SI{2.2}{MHz}$. Red circles, corresponding to the solid line, were taken at $\Omega_p = \Omega_S = 2 \pi \times \SI{2.8}{MHz}$ with both filter cavities. The maximum efficiency achieved in the latter case was 80\%. Error bars denote the standard error of the mean of three repetitions. For the fit results, see Table~\ref{table-fit-results}.}
\label{fig-filter-cavity-effect}
\end{figure}

Finally, we used STIRAP to transfer Feshbach molecules into the absolute ground state $X^1\Sigma^+, v=0$, $|J, m_J, m_{I, \mathrm{Na}}, m_{I, \mathrm{K}}\rangle = |0, 0, 3/2, -4 \rangle$. For this, the lasers were tuned to $\Delta = \delta = 0$ and the previously discussed pulse sequence (see Fig.\ \ref{fig-sequence}) was applied. As seen in Fig.\ \ref{fig-filter-cavity-effect}, we reached transfer efficiencies of up to 80\% when using filter cavities.  Initially, we attempted STIRAP without any filtering and observed a strong dependence of the one-way STIRAP 
efficiency $\eta$ on the pulse duration $\tau$. The one-way efficiency
was also limited to the low value $\eta = 64\%$.

The relatively low efficiency obtained with the unfiltered lasers can be explained with the theoretical framework developed in Refs.~\cite{Yatsenko_2002, Yatsenko_2014} to describe the effect of laser noise on STIRAP efficiency. Following this theory, we expect
\begin{equation}
\label{eq-stirap-efficiency}
\eta = \exp \left(- \frac{\pi^2 \gamma}{\Omega^2 \tau} - \frac{\tau}{4}(D_p + D_S + \Gamma_p + \Gamma_S)\right).
\end{equation}
Here, $D$ stands for the FWHM laser linewidth, and $\Gamma$ denotes an effective linewidth corresponding to the broadband phase-noise pedestal. For both variables, the indices $p$ and $S$ refer to the pump and Stokes lasers, respectively. The first term inside the exponential function corresponds to loss caused by non-perfect adiabaticity of the transfer. The $D$-terms correspond to the noise contributions from the width of the carrier of each laser, and the $\Gamma$-terms to the contributions from broadband noise. For the case of a Lorentzian shape of the broadband noise, $\Gamma$ can be calculated using the formula~\cite{Yatsenko_2014}
\begin{equation}
\label{eq-effective-linewidth}
\Gamma_i = \frac{\Omega_i^2G_i P_i}{4G_i^2 + 2G_i\gamma + \Omega_i^2},
\end{equation}
where $i$ is replaced by $p$ or $S$. Here, $\Omega$ stands for the peak Rabi frequency, $G$ for the FWHM of the laser noise pedestal, and $P$ for the total noise power fraction, i.e., the sum of the phase and intensity noise power fractions of the respective laser. In the following, we assume that $D_p = D_S = 2\pi \times \SI{1}{kHz}$, a typical value likely limited by fiber noise and vibrations of the ULE cavity, which affect the two lasers similarly~\cite{Ma_1994}.

According to Eq.~(\ref{eq-effective-linewidth}), noise at frequencies close to the Rabi frequency is especially detrimental to the transfer efficiency, and this effect becomes stronger at long pulse durations. This problem can be partially mitigated by increasing Rabi frequencies or choosing smaller values of $\tau$. However the line-broadening caused by this can cause other problems such as unwanted addressing of off-resonant hyperfine levels~\cite{Liu_2019b}. Therefore, a reduction of the servo bumps should be a more effective method to improve STIRAP efficiency. We tested this hypothesis by adding filter cavities as described in section~\ref{section-setup}, which indeed led to an improvement to $\eta = 80\%$. In addition, the optimal value of $\tau$ was increased from $\SI{15}{\mu s}$ to $\SI{25}{\mu s}$. By fitting Eq.\ (\ref{eq-stirap-efficiency}) to our data, using the known peak Rabi frequencies and excited state linewidth, we determined $\Gamma_{\mathrm{tot}} = \Gamma_p + \Gamma_S$. The results are shown in Table \ref{table-fit-results}.

\begin{table}
	\caption{Results of the fits of Eq.\ (\ref{eq-stirap-efficiency}) to the data in Fig.\ \ref{fig-filter-cavity-effect}. The values for $P_{\mathrm{tot}}$ are determined using Eq.\ (\ref{eq-effective-linewidth}).}
	\label{table-fit-results}
	\renewcommand{\arraystretch}{1.3}
	\begin{tabular}{l l l l}
		\hline
		Filter cavities \hspace{2mm} & $\Omega_p = \Omega_S$ \hspace{8mm} & $\Gamma_{\mathrm{tot}}$ \hspace{14mm} & $P_{\mathrm{tot}}$ \\ \hline
		None & $2 \pi \times \SI{2.2}{MHz}$ & $2 \pi \times \SI{7.2}{kHz}$ & $4.6 \times 10^{-2}$ \\
		Only pump & $2 \pi \times \SI{2.2}{MHz}$ & $2 \pi \times \SI{2.5}{kHz}$ & $1.6 \times 10^{-2}$ \\
		Both & $2 \pi \times \SI{2.2}{MHz}$ & $2 \pi \times \SI{1.0}{kHz}$ & $6.5 \times 10^{-3}$ \\
		Both & $2 \pi \times \SI{2.8}{MHz}$ & $2 \pi \times \SI{1.6}{kHz}$ & $6.7 \times 10^{-3}$ \\
		\hline
	\end{tabular}
\end{table}

Next, we compared our measurements with the prediction of Eq.~(\ref{eq-effective-linewidth}). As the distribution of noise is not Lorentzian in our case, the model is not expected to give a quantitatively correct description; however we found it to be a simple way to obtain reasonable estimates about the impact of noise on STIRAP. For the comparison, we used the known values of Rabi frequency and excited-state linewidth. We also made the simplifying assumption that $G = 2\pi \times \SI{1.2}{\mega\hertz}$, because the noise power decays very quickly beyond the servo bumps. From this and the fitted values of $\Gamma_\mathrm{tot}$, we obtained the total noise power fraction $P_{\mathrm{tot}} = P_p + P_S$ as shown in Table \ref{table-fit-results}. For each laser without filter cavity, we expect a phase-noise power fraction $P_\phi = 6 \times 10^{-3}$ as well as a contribution from intensity noise which is of similar magnitude. The fitted value of $P_{\mathrm{tot}} = 4.6 \times 10^{-2}$ thus agrees reasonably well with the model for the case without filter cavities. However, we find that the filter cavities appear to provide less than the expected \SI{22}{dB} reduction of total phase noise power. This is likely because they turn some phase noise into intensity noise rather than fully remove it. Another possibility is that $D_p$ or $D_S$ are larger than $2\pi \times \SI{1}{kHz}$, leading to an underestimation of the effect of the filter cavities. In this case, fiber-noise cancellation~\cite{Ma_1994} would be the most effective way to further increase STIRAP efficiency. In future work, it should be possible to obtain quantitative agreement between experiment and theory by measuring the linewidth and the intensity-noise spectrum of the lasers, and by generalizing the theory to the case of non-Lorentzian noise. 

\section{Conclusion}
We have demonstrated that, by using magnetoassociation on the Feshbach resonance at $B=\SI{78.3}{G}$, the closed-channel fraction of Feshbach molecules of $^{23}$Na$^{40}$K can be increased to up to 90\%. This leads to increased pump transition strengths for STIRAP to the absolute ground state and allows us to achieve equal or better transfer efficiencies than previously demonstrated with a simpler laser setup. We have also shown that the use of filter cavities is an effective method to reduce detrimental laser noise and to increase STIRAP efficiency. The analytical model developed by Yatsenko \textit{et al.~}\cite{Yatsenko_2002, Yatsenko_2014} offers a simple and effective way to estimate the improvement that can be achieved by noise reduction. In future work, it may be possible to further improve the transfer efficiency by increasing the Rabi frequencies and reducing the intensity noise of the filter-cavity transmission. In the case of RbCs and NaRb, molecules with a narrow linewidth intermediate state, where Feshbach states with a relatively  large closed-channel fraction have previously been employed, conversion efficiencies of 90\% or better have been demonstrated with low Rabi frequencies. We therefore expect that noise reduction, for example with filter cavities, in a STIRAP scheme with a narrow-linewidth intermediate state will allow reaching near-unit efficiencies. This is an important requirement for high-fidelity state preparation and detection for future molecule quantum gas microscopes as well as quantum computation with molecule arrays.

\begin{acknowledgments}
We thank Eberhard Tiemann for providing a coupled-channel calculation of the Feshbach resonance and Fengdong Jia for his help with the cavity vacuum setup. We also thank Xiangliang Li and Fabian Schmid for insightful discussions. We gratefully acknowledge support from the Max Planck Society, the European Union (PASQuanS Grant No. 817482) and the Deutsche Forschungsgemeinschaft under grants No. 390814868 and FOR 2247.
\end{acknowledgments}

\bibliography{main-bibliography}

\end{document}